\documentclass[apj]{emulateapj}
\usepackage{lscape}

\newcommand{\hii}         {\mbox{\rm \ion{H}{2}}}

\newcommand{\um}          {$\mu$m}

\newcommand{\msun}        {M$_{\odot}$}
\newcommand{\ha}          {\mbox{H$\alpha$}}

\newcommand{\hr}          {\mbox{$^h$}}
\renewcommand{\min}       {\mbox{$^m$}}
\renewcommand{\sec}       {\mbox{$^s$}}

\newcommand{\bracket}   {[3.6]}
\newcommand{\simgtr}{\, \raisebox{-.2ex}{$\stackrel{>}{\mbox{\tiny $\sim$}}$} \,}

\shorttitle{S$^{3}$MC: Discovery of Embedded Protostars in NGC 346}
\shortauthors{Simon et al.}  
\slugcomment{Accepted for publication in \emph{The Astrophysical Journal}}

\begin{document}

\title{The \emph{Spitzer} Survey of the Small Magellanic Cloud:
  Discovery of Embedded Protostars in the \hii\ region NGC 346}

\author{Joshua D. Simon\altaffilmark{1}, Alberto D.
  Bolatto\altaffilmark{2}, Barbara A. Whitney\altaffilmark{3}, Thomas
  P. Robitaille\altaffilmark{4}, Ronak Y. Shah\altaffilmark{5}, David
  Makovoz\altaffilmark{6}, Sne\v{z}ana
  Stanimirovi\'{c}\altaffilmark{7}, Rodolfo H. Barb{\'
    a}\altaffilmark{8}, and M{\' o}nica Rubio\altaffilmark{9}}

\altaffiltext{1}{Department of Astronomy, California Institute of Technology,
                 1200 E. California Blvd, MS 105-24, Pasadena, CA  91125;
                 jsimon@astro.caltech.edu}

\altaffiltext{2}{Department of Astronomy, University of California at
                 Berkeley, 601 Campbell Hall, Berkeley, CA 94720;
                 bolatto@astro.berkeley.edu}

\altaffiltext{3}{Space Science Institute,
                4750 Walnut Street, Suite 205, Boulder, CO 80301;
	        bwhitney@spacescience.org}

\altaffiltext{4}{SUPA, School of Physics and Astronomy, University of 
                St. Andrews, North Haugh, St. Andrews, KY16 9SS, UK;
	        tr9@st-andrews.ac.uk}

\altaffiltext{5}{Institute for Astrophysical Research, Boston University, 
                725 Commonwealth Avenue, Boston, MA 02215;
	        ronak@bu.edu}

\altaffiltext{6}{\emph{Spitzer} Science Center, California Institute
                of Technology, 1200 E. California Blvd., 
                MS 220-6, Pasadena, CA 91125; 
                davidm@ipac.caltech.edu}

\altaffiltext{7}{Department of Astronomy, University of Wisconsin,
                475 North Charter Street, Madison, WI  53706;
	        sstanimi@astro.wisc.edu}

\altaffiltext{8}{Departamento de F{\' i}sica, Universidad de La Serena,
                Benavente 980, La Serena, Chile;
	        rbarba@xeneize.dfuls.cl}

\altaffiltext{9}{Departamento de Astronom{\' i}a, Universidad de Chile,
                Casilla 36-D, Santiago, Chile;
	        monica@das.uchile.cl}

\begin{abstract}

We use \emph{Spitzer Space Telescope} observations from the
\emph{Spitzer} Survey of the Small Magellanic Cloud (S${^3}$MC) to
study the young stellar content of N66, the largest and brightest
\hii\ region in the SMC.  In addition to large numbers of normal
stars, we detect a significant population of bright, red infrared
sources that we identify as likely to be young stellar objects (YSOs).
We use spectral energy distribution (SED) fits to classify objects as
ordinary (main sequence or red giant) stars, asymptotic giant branch
stars, background galaxies, and YSOs.  This represents the first
large-scale attempt at blind source classification based on
\emph{Spitzer} SEDs in another galaxy.  We firmly identify at least 61
YSOs, with another 50 probable YSOs; only one embedded protostar in
the SMC was reported in the literature prior to the S${^3}$MC.  We
present color selection criteria that can be used to identify a
relatively clean sample of YSOs with IRAC photometry.  Our fitted SEDs
indicate that the infrared-bright YSOs in N66 have stellar masses
ranging from 2~\msun\ to 17~\msun, and that approximately half of the
objects are Stage II protostars, with the remaining YSOs roughly
evenly divided between Stage I and Stage III sources.  We find
evidence for primordial mass segregation in the \hii\ region, with the
most massive YSOs being preferentially closer to the center than
lower-mass objects.  Despite the low metallicity and dust content of
the SMC, the observable properties of the YSOs appear consistent with
those in the Milky Way.  Although the YSOs are heavily concentrated
within the optically bright central region of N66, there is ongoing
star formation throughout the complex and we place a lower limit on
the star formation rate of $3.2 \times 10^{-3}$~\msun~yr$^{-1}$ over
the last $\sim$1~Myr.
\end{abstract}

\keywords{Magellanic Clouds --- \hii\ regions --- infrared: stars ---
  ISM: individual (N66) --- stars: formation --- stars: pre-main
  sequence}

\section{INTRODUCTION}

Star formation is one of the most important astrophysical processes,
but because it takes place on small physical scales and behind heavy
optical obscuration, observations of the early phases of star
formation have generally been limited to the Milky Way.  Only in
recent years have these studies begun to be extended to extragalactic
objects.  Now, the high sensitivity, high angular resolution, and
large field of view provided by the \emph{Spitzer Space Telescope} at
mid-infrared wavelengths open a new window on recently formed stars in
nearby galaxies like the Magellanic Clouds.  Studying these young
stellar objects (YSOs) provides an excellent complement to
observations of star formation in the Milky Way because the physical
conditions they are subject to are quite different from those in the
Galaxy.  In particular, the metallicity of the Magellanic Clouds is
well below solar, and their dust content is correspondingly lower as
well \citep{snez00,leroy07}.  Most star formation today is taking
place in galaxies smaller than the Milky Way (for which the Magellanic
Clouds are good prototypes), and star formation at high redshift
occurred in low-metallicity environments, so investigating the effects
of these differences on how star formation works will be an important
step towards understanding how many of the stars in the universe were
formed.

The Small Magellanic Cloud (SMC), and its more massive companion the
Large Magellanic Cloud (LMC), are the two nearest star-forming dwarf
galaxies.  Located at a distance of 61.1~kpc \citep*{westerlund,storm04,
  hhh05,kw06}, and with a luminosity of $\sim6 \times 10^{8}$~L$_{\odot}$
\citep{rc3} and a metallicity of $\sim1/5$ solar
\citep*{dufour75,ptp76,dh77,ppr00}, the SMC is perhaps the best local
analog of primitive galaxies.  The SMC is actively forming stars at a
rate of $\sim0.05$~M$_{\odot}$~yr$^{-1}$ \citep{wilke04}, and is
populated by well studied \hii\ regions and young star clusters.  The
most active star-forming region is NGC~346 \citep[alternately known as
  N66;][]{henize}, located toward the northern end of the SMC bar.
The OB association powering N66 contains 33 spectroscopically
confirmed O stars, and a similar number of bright blue stars without
spectra that are likely to be O stars as well \citep*{mpg89}

In this paper, we employ \emph{Spitzer} observations from the
\emph{Spitzer} Survey of the Small Magellanic Cloud (S$^{3}$MC;
\citealt{bolatto07}, hereafter B07) to locate objects in N66 with
mid-infrared excesses that we identify as candidate YSOs.  In the
following section, we very briefly describe the observations and our
photometry.  In \S \ref{results}, we classify the detected sources by
fitting their spectral energy distributions and study their locations
in color-color space.  In \S \ref{discussion} we discuss some of the
implications of our results and compare to recent optical studies of
N66.  We summarize our findings in \S \ref{conclusions}.

\section{Observations, Data Reduction, and Analysis}
\label{observations}

The S$^{3}$MC is a project to map the star-forming body of the SMC
with \emph{Spitzer} in all seven Infrared Array Camera
\citep[IRAC;][]{irac} and Multiband Imaging Photometer for
\emph{Spitzer} \citep[MIPS;][]{mips} bands.  The images cover an area
of $\sim3$~deg$^{2}$ including the entire bar and wing of the SMC,
with an average exposure time of 144~s at each position.  The MIPS
data were obtained in 2004 November and the IRAC data in 2005 May.
The data were processed with version S11.4 of the automated
\emph{Spitzer} pipeline.  We constructed mosaic images from the
individual Basic Calibrated Data frames using the Mosaicking and Point
Source Extraction (MOPEX) software provided by the \emph{Spitzer}
Science Center (SSC).  B07 describe further details of the
observations and the data processing.

We performed photometry on the mosaic images with the Astronomical
Point Source Extraction (APEX) tasks in the MOPEX package
\citep{apex}.  We selected a set of $20-30$ bright stars in each band
that were as isolated as possible and constructed point response
functions (PRFs) directly from the data.  We then fit these PRFs to
every detected source in the images to determine fluxes.  The images
contain extensive diffuse emission that must be separated from point
sources, so we used a small median filter ($8\farcs4$) to remove the
background, and then detected sources on the background-subtracted
image.  Because the median filtering removes some flux even from point
sources, the PSF fitting then took place on the original (not
background subtracted) image.  We found that this technique offered
the best compromise between detecting bright sources on top of diffuse
emission and detecting faint sources in background-free regions.  For
the 24~\um\ MIPS data, in which the extended emission dominates over
the point sources, we used a somewhat larger median filter
($27\farcs5$) and switched to the `combo' algorithm for separating
clusters of bright emission into individual sources in place of the
`peak' algorithm in the task {\sc detect}.\footnote{More information
  about the {\sc detect} task and its associated options can be found
  in the documents titled \emph{APEX User's Guide} and \emph{Image
    Segmentation} (both by D. Makovoz) that are available on the
  \emph{Spitzer Science Center} website.}  We used observations of
bright, isolated point sources in the SMC to measure the aperture
corrections for the photometry.  To provide near-infrared fluxes, we
used VLT/ISAAC J and K$_{s}$ imaging of the center of N66
\citep[][Rubio \& Barb\'{a}, in preparation]{rubio02} and Two Micron
All Sky Survey \citep[2MASS;][]{2mass} measurements in the outer
regions.  We also added \emph{HST} $V$- and $I$-band photometry of the
central region from \citet{gouliermis06}, but because the much higher
angular resolution of the ACS images can cause confusion within the
\emph{Spitzer} resolution element, we only used these data for bright
stars ($V \le 17.5$).

The assumed photometric uncertainty associated with each flux
measurement is very important for determining the relative weights
given to the various data points in the SED fitting that we carry out
in \S \ref{sedfitting}.  Although the statistical uncertainties on all
of the photometric measurements were quite small (generally a few
percent or less), we imposed larger minimum uncertainties on all of
the data points to account for systematics.  For example, the
\emph{HST}, 2MASS, VLT, IRAC, and MIPS observations were made at
various times over a period of 7 years, providing plenty of time for
the fluxes of the YSOs, which may be variable, to change.  The
absolute photometric accuracy of IRAC is 10\% \citep{irac}, so we
added 10\% of the flux in quadrature to the measured uncertainty for
each source.  MIPS also has a stated accuracy of 10\% \citep{mips}, so
when combined with the IRAC calibration accuracy and the 6 month time
baseline between the observations, we used a 20\% minimum error for
MIPS fluxes.  Since the near-IR observations took place $6-7$ years
earlier, we imposed larger minimum errors of 35\% (for the deep VLT
data) and 50\% (for the shallow 2MASS data) on those data points.  The
optical measurements are more recent (2004 July) and more accurate, so
we assumed 10\% uncertainties on the \emph{HST} photometry.

\section{RESULTS}
\label{results}

\subsection{Photometry}
\label{photometry}

For the purposes of this study, we limited our point-source fitting to
a $12\farcm75 \times 12\farcm75$ region covering the full extent of
the hot dust emission from N66 seen in our 24~\um\ image (from
00\hr57\min51\sec\ to 01\hr00\min37\sec\ and from
$-$72\degr16\arcmin48\arcsec\ to $-$72\degr04\arcmin19\arcsec\ [all
  coordinates J2000.0]).  In this box, we detect 6544 sources at
3.6~\um, 5836 at 4.5~\um, 1784 at 5.8~\um, 1718 at 8.0~\um, 101 at
24~\um, and 15 at 70~\um, for a total of 8011 unique objects.  A
significant number of these objects have \emph{Spitzer} colors that
are redder than normal stars should be at these wavelengths,
suggesting that they may be YSOs or background galaxies (B07).

\subsection{SED Fitting}
\label{sedfitting}

There are two possible approaches to determining the nature of
individual \emph{Spitzer} sources.  One could simply use the observed
colors and magnitudes to classify the sources into various categories.
Stars of nearly all kinds (except those with dusty atmospheres) have
colors near zero for any combination of \emph{Spitzer} bands because
the IRAC and MIPS bands are on the Rayleigh-Jeans tail of their
spectral energy distributions (SEDs).\footnote{Note that we use the
  Vega magnitude system throughout this paper.}  YSOs, because of the
emission from warm ($T \sim 200$~K) dust around the central protostar,
have red colors throughout the mid-IR.  IRAC colors for theoretical
YSO models are given by, e.g., \citet{allen04} and
\citet{whitney03a,whitney03b,whitney04b}.  Alternatively, one can
compare the full SEDs to a variety of source models and find the best
match for each object.

The advantage of the color selection strategy is its simplicity, but
it also has some drawbacks.  It fails to use all of the available
information about each object (since we also have MIPS, optical, and
near-IR fluxes for many of the sources), and it is very difficult to
learn about the detailed properties of individual sources from only
their colors.  Color selection also does not offer a way to test the
assumption that SED models designed for Milky Way sources offer a good
description of YSOs that form in the metal- and dust-poor environment
of the SMC; this is a disadvantage for studies seeking to determine
how star formation proceeds at low metallicity, but could also be an
advantage in that it enables the selection of YSOs without regard to
the properties of their environment.  Because the SED classification
should be more accurate, we begin with that technique and then compare
the results to color selection in \S \ref{colors}.

The SED fitting tool used for this study employs a linear regression
method to find all the SEDs from a large grid of models that fit the
data within a specified $\chi^{2}$ \citep{robitaille07}.  The grid of
models consists of 7853 stellar atmospheres \citep{kurucz92,bh05}
encompassing all available metallicities and effective temperatures, a
limited number of \emph{Infrared Space Observatory} (ISO) spectra of
galaxies \citep{dale05,silva98} and AGB stars
\citep{sylvester99,olivier01,hony02a,molster02,hony02b,fn03}, and the
20,000 YSO models from \citet{robitaille06} computed using the
radiation transfer codes from
\citet{whitney03a,whitney03b,whitney04b}.  Each YSO model outputs SEDs
for 10 viewing angles, so the YSO grid effectively contains $2 \times
10^{5}$ SEDs.  The foreground extinction, $A_V$, is fit simultaneously
using an extinction law derived from GLIMPSE observations (Indebetouw
et al. 2006).  At ultraviolet and visible wavelengths, this extinction
law is not appropriate for the SMC, but in the near-IR and mid-IR the
differences in extinction between the SMC and Galaxy are small
\citep{gordon03,cartledge05}.  The fitter is run first using only the
stellar atmosphere grid.  It is then run three more times using the
YSO grid, ISO galaxy spectra, and ISO AGB spectra on the sources that
are not well-fit by stellar atmospheres.  Based on all the successful
fit results, defined by their $\chi^{2}$ values (see below), we can
classify sources and calculate best estimates and uncertainties for
each model parameter.

We fit an SED for every object for which we had at least four flux
measurements.  To better constrain the fits at long wavelengths where
most of the sources were not detected, we added 24~\um\ upper limits
of 1~mJy (5 times the limiting sensitivity of the 24~\um\ data) for
each source that was not detected at 24~\um.  Because there are many
more model parameters than data points that go into each observed SED,
the traditional reduced $\chi^{2}$ statistic cannot be used to
evaluate the success of the fits.  As a proxy, we calculate the
$\chi^{2}$ per data point (where each SED has between 4 and 11 data
points).  Given this definition, it is difficult to establish \emph{a
  priori} the division between good and poor fits.  We inspected the
fits by eye and determined that fits with a $\chi^{2}$ per data point
of less than 2.2 could reasonably be interpreted as good fits, while
higher $\chi^{2}$ values indicated fits that were not a good match to
the templates.  We will therefore use this cutoff point to separate
good and poor fits throughout the paper.  Note that this definition
depends on the flux uncertainties imposed in \S \ref{observations}.

\subsubsection{Results From Successful SED Fits}

\begin{figure}[t!]
\epsscale{1.20}
\plotone{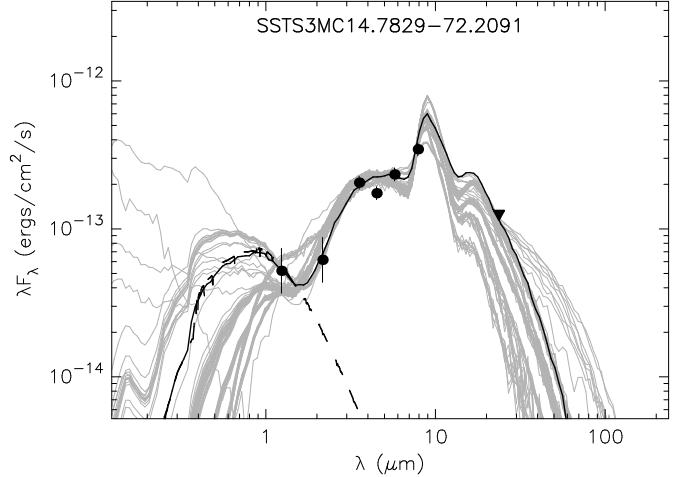}
\caption{Sample SED fits for a YSO in N66.  The black points indicate
  the measured fluxes and uncertainties, and the black triangles are
  for upper limits.  The thin black line represents the best fitting
  SED, and the gray lines represent all other acceptable
  ($\chi^{2}$/data point $\le 2.2$) YSO fits.  The dashed line
  represents the stellar photosphere model (including the effect of
  foreground extinction).  This object is a Stage II YSO with a
  stellar mass of 4.9~\msun\ (see \S \ref{stages} for the definition
  of YSO stages).}
\label{fits}
\end{figure}

Out of the 1645 sources with four or more fluxes, 1322 (80.4\%) are
succesfully fit with stellar SEDs (see above for what we mean by a
``successful fit'').  Of the remaining 323 sources, 61 can only be fit
successfully by YSO models, 27 are background galaxies, 6 are AGB
stars, 81 can be fit by multiple classes of models, and 148 cannot be
fit by any of the available templates.  We display one example YSO SED
in Figure \ref{fits}.

For the sources that can be well-described by multiple types of
templates, we used the $\chi^{2}$ value of the best fit for each
template type to attempt a more accurate classification.  We set a
cutoff of 1.5 times the best $\chi^{2}$ value, and if, for example,
the best YSO fit had $\chi^{2} = 1.0$ and the best galaxy fit had
$\chi^{2} = 1.6$ then we classified the object as a probable YSO.
This choice of a $\chi^{2}$ cutoff indeed results in the majority of
the uncertain objects being classified as YSOs and leaves very few
objects with unknown types, which appears to be a reasonable outcome.
While the exact value chosen for the cutoff is arbitrary, it is clear
from the spatial distribution of the various object types that many of
the sources classified as possible or definite galaxies must actually
be YSOs.  The results of this paper do not depend strongly on the
exact value of the cutoff.  Using this method, we were able to obtain
probable classifications for 68 of the 81 objects of uncertain type,
including 50 YSOs, 13 galaxies, and 5 AGB stars.  Thus, the final
object counts are 111 YSOs, 40 galaxies, and 11 AGB stars.  The
remaining 13 objects had $\chi^{2}$ values for multiple object types
that were too close to distinguish reliably.  We list selected
parameters from the fits for all 111 YSOs in Table \ref{ysotable}, and
we summarize the source classification results in Table
\ref{sourcetypes}.

\subsubsection{Objects Poorly Fit by the SED Models}

We inspected each of the 148 poorly fit sources and the various
attempted fits to their SEDs to determine why the fitting failed for
them.  We found that slightly more than 1/3 of the objects had SEDs at
short wavelengths that appeared stellar in origin, but the long
wavelength data (often at 24~\um) were significantly in excess of the
extrapolated photospheric emission.  The fits to these sources appear
to have failed for a number of reasons including slight mismatches
between the optical and IR photometry (either from variability or
measurement error), confusion, misidentification of the long
wavelength counterparts, and contamination of the photometry by
underlying dust emission (revealed by PAH features).  However, some of
these sources may also have true infrared excesses indicating the
presence of circumstellar material.  Another $\sim1/3$ of the bad fits
were generally faint sources that are spatially coincident with dust
filaments in the \emph{Spitzer} images, again causing significant
contamination of their SEDs by PAH features in the IRAC bands.  Some
of these objects are probably point sources (either stars or YSOs),
while others may simply be unresolved dust knots.  The final 1/3 of
the bad fits was a mixed population whose SEDs could not be
straightforwardly interpreted.  Photometric errors in one or more
bands and blending were likely responsible for the failure to fit
these objects, but they represented only 2.6\% of the total sample.

\subsection{IRAC Colors of YSOs}
\label{colors}

\begin{figure*}[t!]
\epsscale{1.20}
\plotone{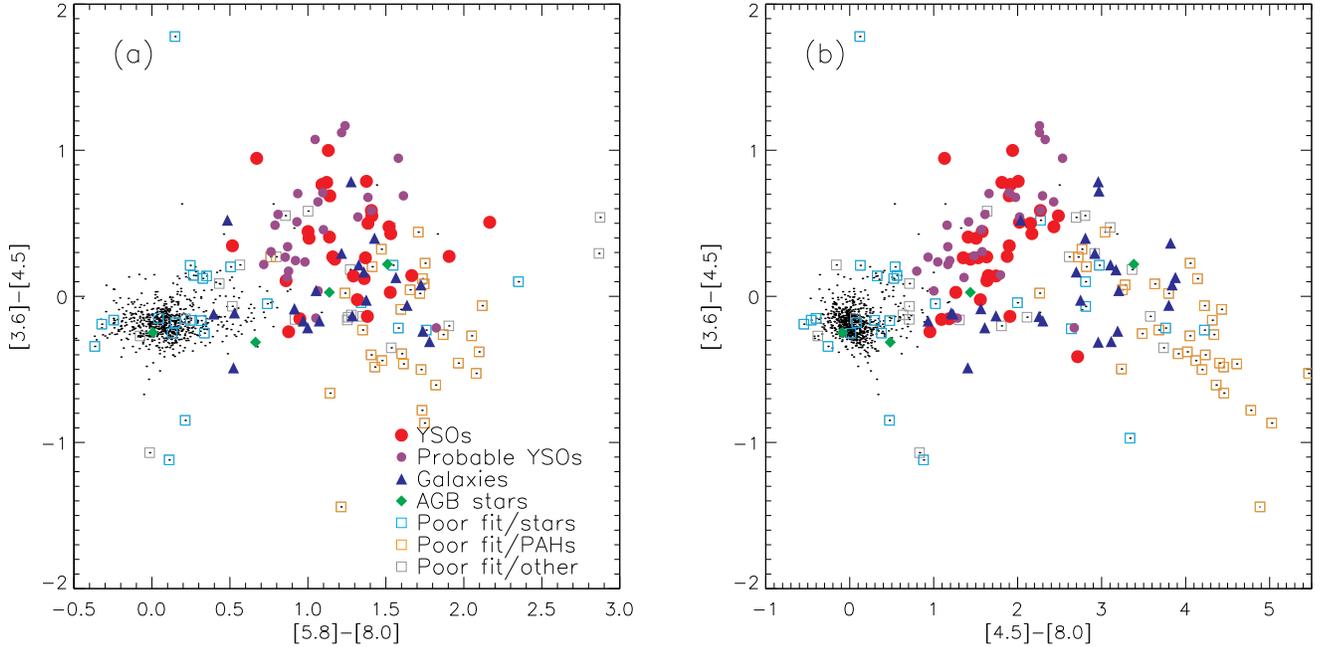}
\caption{(\emph{a}) IRAC [3.6]$ - $[4.5] vs. [5.8]$ - $[8.0]
  color-color diagram of N66.  The black dots represent all detected
  sources (primarily stars), the filled red circles represent objects
  that can only be fit by YSO models, the smaller purple circles
  represent objects that are best fit by YSO models, although galaxy
  and/or AGB fits with significantly higher $\chi^{2}$ values are also
  acceptable, the filled blue triangles are background galaxies, and
  the filled green diamonds are AGB stars.  The open squares represent
  sources for which satisfactory fits were not obtained: the cyan
  symbols have SEDs that suggest they are likely to be stars, the
  orange symbols are sources that are contaminated by PAH emission
  features, and the gray symbols are the remaining unclassified poor
  fits.  Note that the separation of YSOs from the other classes of
  sources is not very clean in this diagram.  (\emph{b}) IRAC [3.6]$ -
  $[4.5] vs. [4.5]$ - $[8.0] color-color diagram of N66.  Symbols are
  the same as in (\emph{a}).  This plot offers the best separation of
  YSOs from the various contaminants, although there are still a few
  extragalactic sources that have similar colors to the YSOs.}
\label{ysocolors}
\end{figure*}

Using the results of the SED fitting, we can now investigate the
colors of the objects classified as YSOs and as stars.  We display
four color-color plots in Figure \ref{ysocolors} to illustrate the
possibilities for color selection.  Stars, as expected, lie in the
cloud of points centered near (0,~0), while sources that are red in
one or both colors may be YSOs.  Note that the frequently used [3.6]$
- $[4.5] vs. [5.8]$ - $[8.0] color-color plot (Fig. \ref{ysocolors}a)
does not cleanly separate YSOs from other types of sources,
particularly stars with modest IR excesses and sources with PAH
contamination.  Better separation can be achieved using different
combinations of colors, as shown in Figure \ref{ysocolors}b.  These
plots take advantage of longer color baselines and the abrupt change
in YSO spectra between the 4.5~\um\ and 5.8~\um\ bands to distinguish
YSOs from stars, galaxies, and PAHs.  These results demonstrate that
while color selection can be a useful technique for identifying YSOs,
it does not appear possible to obtain a YSO sample that is both
complete and clean with a simple set of color diagnostics.  SED
fitting is a more comprehensive way to determine the nature of sources
if measurements in enough bands are available.  Nevertheless, because
in many cases it is desirable to classify objects with easily
applicable techniques, we use our data to provide guidance for color
selection of YSOs.

Assuming that the purple points in Figure \ref{ysocolors} are indeed
YSOs, one can select YSOs with the following set of criteria:

\begin{eqnarray}
\nonumber [3.6] - [4.5] & > & 0.6\times([4.5] - [8.0]) - 1.0 \\
\nonumber [4.5] - [8.0] & < & 2.8 \\
\nonumber [3.6] - [4.5] & < & 0.6\times([4.5] - [8.0]) + 0.3 \\
\bracket - [4.5] & > & -([4.5] - [8.0]) + 0.85.
\label{colorselection1}
\end{eqnarray}

\noindent These criteria were defined to maximize completeness; one
could alternatively choose to minimize contamination at the cost of
increased incompleteness, but given the distribution of sources in
Figure \ref{ysocolors}\emph{b}, the differences would be small.  This
selection is only 7\% incomplete for YSOs in our sample that have
measured fluxes at 3.6~\um, 4.5~\um, and 8.0~\um, and has a
contamination of less than 27\%\ (the majority of the objects in this
color box with uncertain classifications may still be YSOs).
Approximately equivalent results can be obtained by substituting the
[3.6]$-$[8.0] color for the [4.5]$-$[8.0] color, but it is slightly
more difficult to avoid picking up stars on the blue end of the
[3.6]$-$[8.0] color axis.

\subsubsection{Comparison to B07 Photometric Selection}

B07 identified a set of very bright YSOs across the entire SMC based
on their 8.0~\um\ magnitudes and [5.8]$-$[8.0] colors.  If we select
the same region of the color-magnitude diagram in N66, we find 15
sources, only 3 of which are conclusively classified as YSOs.  Almost
all of the remaining objects have SEDs that are not well-fit by any of
the models we apply, primarily because of PAH contamination.  The PAH
contamination across the rest of the galaxy should be much lower than
in N66, so these numbers clearly represent a lower limit to the
fraction of actual YSOs in the B07 sample.

\subsection{YSO Masses and Stages}
\label{stages}

YSOs are traditionally divided into classes based on their observed
spectral indices as originally defined by \citet{lada87}.  Because
spectral indices can vary with inclination angle as well as with
evolutionary state, \citet{robitaille06} described their YSO models in
terms of ``stages'', which are analogous to the usual classes, but are
based on the physical quantities that define the evolutionary stage of
the models.  When comparing these models to data (rather than
considering only observations), it therefore makes sense to use the
stage system.  The definitions of each stage rely on the ratio of the
disk mass and envelope accretion rate to the central stellar mass.
Stage I sources have $\dot{M}/M_{*} > 10^{-6}$, Stage II sources have
$\dot{M}/M_{*} < 10^{-6}$ and $M_{disk}/M_{*} > 10^{-6}$, and Stage
III sources have $\dot{M}/M_{*} < 10^{-6}$ and $M_{disk}/M_{*} <
10^{-6}$, where $\dot{M}$ is the envelope accretion rate.  Note that
we assumed a dust-to-gas ratio of $2 \times 10^{-3}$ (1/5 of the Milky
Way value) for N66 \citep{leroy07}.\footnote{The dust-to-gas ratio
  does not affect the YSO fits themselves, but does change the mass
  ratios between the disk and the central star, which is used for the
  stage classification.}

Using this classification scheme, we have grouped the 111 definite and
probable YSOs in N66 into stages.  The SEDs of most objects can be fit
by more than 1 YSO model, so we first selected the models that produce
a $\chi^{2}$ that is within 1 of the best $\chi^{2}$ for each object.
We then computed a stage for each of the acceptable models and
calculated a weighted average stage, using the $\chi^{2}$ values as
weights.  The averages were rounded to the nearest integer to produce
a classification.  We find that of the 111 YSOs, 33 are Stage I, 50
are Stage II, and 28 are Stage III.

We also calculated YSO masses with the same weighted averaging scheme,
and defined the uncertainty on the mass to be the weighted standard
deviation of the masses of the acceptable models.  We found YSOs with
a range of masses from $2.4-16.6$~M$_{\odot}$, including 19 objects
that appear to be proto-OB stars ($M \ge 8$~M$_{\odot}$).  We list the
names, positions, luminosities, luminosity uncertainties ($\Delta$L,
the weighted standard deviation of the luminosities of the acceptable
models), masses, mass uncertainties ($\Delta$M, as defined in the
first sentence of this paragraph), stages, and stage uncertainties
($\Delta$Stage, the weighted standard deviation of the stages of the
acceptable models) of all of the YSOs in Table \ref{ysotable}.

\subsection{Spatial Distribution of YSO Candidates}
\label{distribution}

\begin{figure}[t!]
\epsscale{1.20}
\plotone{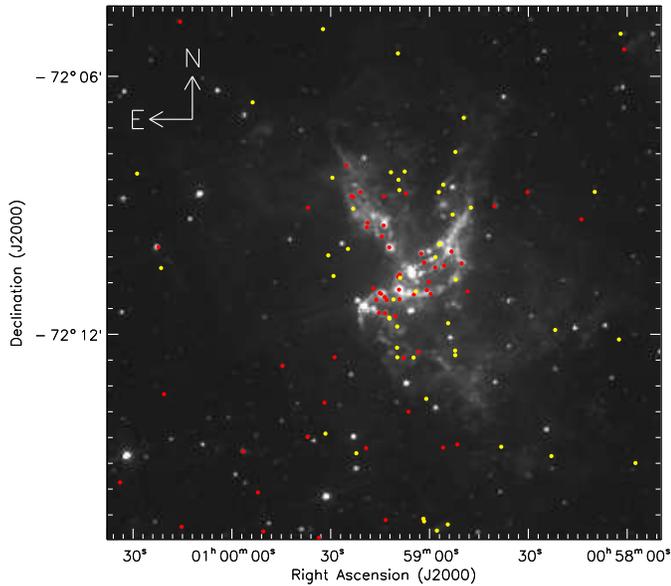}
\caption{Spatial distribution of YSOs in N66, overplotted on an image
  of the 8.0~\um\ emission.  The red circles represent objects with
  SEDs that can only be fit by YSO models, and the yellow circles
  represent objects that are probably YSOs but can be fit by other
  source types as well (albeit with significantly higher $\chi^{2}$
  values).  The image is displayed on a logarithmic scale from
  4~MJy~sr$^{-1}$ to 20~MJy~sr$^{-1}$. }
\label{spatialdist}
\end{figure}

In Figure \ref{spatialdist} we plot the spatial distribution of the
YSOs in N66.  The protostars are obviously highly concentrated towards
the peaks of the 8.0~\um\ emission.  However, there is also star
formation taking place outside of the dense dust cloud that marks the
optical \hii\ region, particularly to the south and southeast.
\citet{mpg89} hypothesized that star formation in N66 has proceeded
from the southwest to the center of the present-day \hii\ region, but
we do find a few probable YSOs southwest of N66, indicating that at
least modest star formation has taken place there within the past few
million years.

\begin{figure}[t!]  
\epsscale{1.20}
\plotone{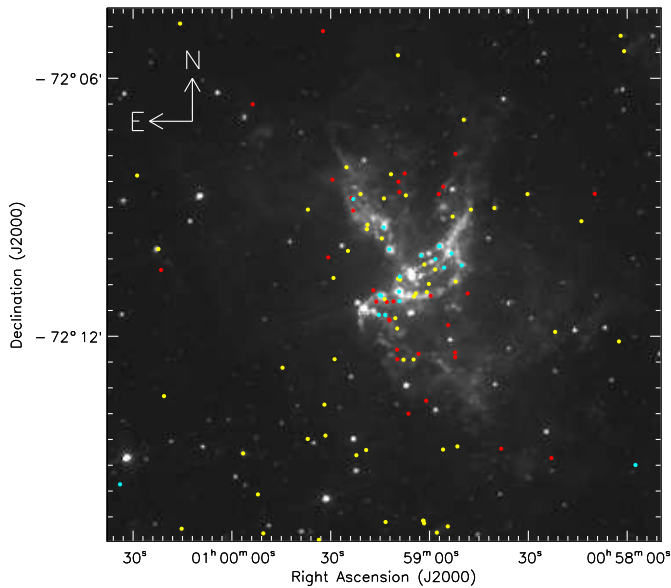} 
\caption{Mass distribution of YSOs in N66, overplotted on an image of
  the 8.0~\um\ emission.  The most massive sources ($M > 8$~\msun) are
  plotted as cyan points, and then decreasing masses are indicated by
  yellow (4.5~\msun$ < M \le 8$~\msun) and red ($M \le 4.5$~\msun)
  points.  The high-mass stars are more concentrated in the center of
  the \hii\ region than the less massive objects.  Note that the
  objects shown in this figure include the full YSO sample, not just
  the 61 definite YSOs that can only be fit by YSO SEDs.}
\label{massdist} 
\end{figure}

The distribution of YSOs throughout the \hii\ region as a function of
mass and stage is not uniform.  We find that the most-embedded objects
(Stage I) are slightly more concentrated towards the center of the
\hii\ region than the more advanced (and presumably older) YSOs.  We
also see evidence for mass segregation, with the most massive objects
exhibiting a strong preference for locations close to the center (see
Figure \ref{massdist}).  All but 2 of the YSOs with $M \ge
8$~M$_{\odot}$ lie on top of bright dust filaments in the main
\hii\ region, and many of them are coincident with molecular peaks and
optical star clusters \citep{rubio00,sabbi07}.  If these massive YSOs
are indeed single objects rather than multiple unresolved sources,
then the mass segregation must be primordial in origin, as the YSOs
have not had time to move very far from their birthplaces.

\section{DISCUSSION}
\label{discussion}

The census of previously known embedded protostars in the SMC consists
of a single object discovered by \citet{gatley82}.  \citet{beaulieu01}
and \citet{dewit03} used variability data and \ha\ imaging to uncover
a small sample of Herbig Ae/Be stars in the SMC, and \citet{nota06}
discovered several hundred low-mass pre-main-sequence stars in N66 via
isochrone fits to \emph{HST} photometry.  These objects, however, are
generally much more evolved than the YSOs identified by this study,
many of which are still embedded in their natal dust clouds and are
therefore faint or invisible at optical wavelengths.  The distribution
of YSOs is concentrated towards clumps of molecular gas traced by the
CO(2-1) emission line \citep{rubio00} and the peaks of the dust
emission at 7~\um\ \citep{contursi00}.  \citet{rubio00} showed that
dense H$_{2}$ knots are associated with these molecular clumps and
suggested that massive star formation could be taking place there.
The large number of embedded YSOs found in this study confirms that
prediction.  Moreover, IR spectroscopy of the 3 brightest embedded
sources detected with ground-based near-IR imaging confirm that these
sources are YSOs (Rubio and Barb{\' a}, in preparation).

The sample of Stage I, II, and III YSOs that we have identified in N66
presents the first opportunity for studying in detail a large sample
of embedded YSOs in another galaxy \citep[see also][]{jones05,chu05},
and more importantly, one whose ISM properties differ substantially
from those of the Milky Way.  The fact that we are able to identify
over 100 YSOs in N66, as well as obtain successful SED fits for almost
all of the other sources, suggests (perhaps surprisingly) that
protostars in the SMC resemble the YSO models constructed by
\citeauthor{whitney03a} for Milky Way objects, even though the
metallicity and dust-to-gas ratio are a factor of $\sim5$ lower in the
SMC.  The only potential difference between Milky Way and SMC YSOs
that is evident in our results is that some of the sources are best
fit by relatively cool photospheres, but the YSO models with such
photospheres do not have high enough luminosities to match their
observed brightnesses at the distance of the SMC.  If these objects
are actually YSOs, this problem could be an indication that accretion
is continuing even after these stars have reached the main sequence,
which leads to expanded photospheres and lower temperatures compared
to normal pre-main sequence tracks and zero-age main-sequence
photospheres \citep{mt03}.  An alternative possibility is that these
sources actually consist of multiple cool protostars, which would
explain their unusually high luminosities.  The luminosities of these
objects are $2-3$ orders of magnitude higher than the expected
luminosities of individual YSOs of the same temperature, however,
which makes the multiplicity explanation appear unlikely.  Finally,
evolved stars in the post-AGB phase also have expanded cool
photospheres, so it is possible that some of the ``cool luminous''
sources could be post-AGB stars.  Additional modeling and observations
of some representative objects may be necessary to resolve this issue.

\subsection{The Star Formation Rate in N66}

\begin{figure}[t!]  
\epsscale{1.20}
\plotone{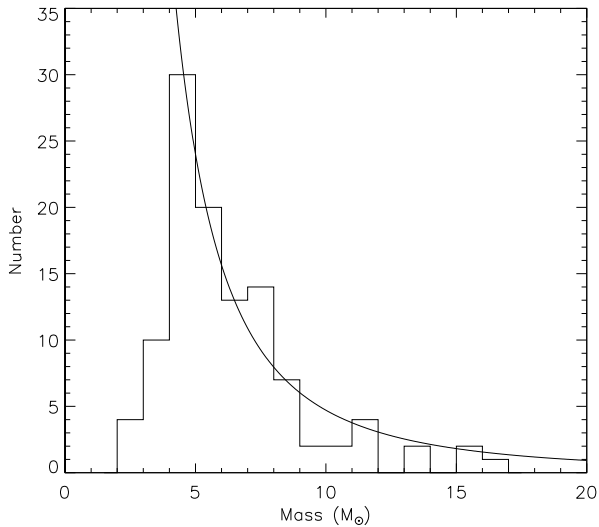} 
\caption{Mass function of YSOs in N66.  The solid curve shows the
  best-fitting Salpeter IMF, which provides a reasonable fit to the
  data above 4~\msun, where the observational incompleteness becomes
  severe.}
\label{imf} 
\end{figure}

The 111 YSOs in our sample have a combined stellar mass of 692~\msun.
The observed mass function (see Figure \ref{imf}) turns over at
$\sim4$~\msun, indicating that incompleteness becomes serious at this
point.  Artificial star tests in our photometry show that we are 90\%
complete even in the confused central region of N66 down to flux
levels of 300~$\mu$Jy, 200$\mu$Jy, 200$\mu$Jy, and 300~$\mu$Jy from
3.6-8.0~\um.  Applying these limits to the full library of YSO models
confirms that our incompleteness is severe below 4~\msun.  If we
assume that star formation in N66 follows a \citet{salpeter} initial
mass function (IMF) down to 0.1~\msun, we calculate that the total
mass in protostars for the entire \hii\ region is $\sim3160$~\msun.
In reality, this is a lower limit to the mass because even at
4~\msun\ the data are somewhat incomplete.  If these YSOs have all
formed within the last $\sim1$~Myr, then the average star formation
rate over that time is $3.2 \times 10^{-3}$~\msun~yr$^{-1}$.  Thus, we
find that N66 comprises at least $\sim6$\% of the total current star
formation in the SMC.

\subsection{Comparison with Optical Observations}

\citet{nota06}, \citet{gouliermis06}, and \citet{sabbi07} used
\emph{HST} imaging to study the young stellar population of N66 in the
optical.  \citet{sabbi07} identified 16 subclusters of pre-main
sequence stars (their Figure 8; several of these clusters were also
pointed out by \citeauthor{gouliermis06}) in the \hii\ region, and we
find that all but two of these also have YSOs associated with them
(see below).  \citeauthor{sabbi07} estimated ages of $\sim3$~Myr for
the first 15 of the subclusters, but our detection of YSOs
demonstrates that star formation has continued until the present day
in these areas.

Subcluster 1 (Sc~1) is nearly coincident with an extremely bright
\emph{Spitzer} source (the 4th-most luminous object in the field at
8.0~\um), SSTS3MC~14.7725-72.1766, which we are able to fit with both
YSO and AGB SEDs with similar $\chi^{2}$ values.  Given the location
of this object at the center of the NGC~346 cluster, surrounded by
numerous very young massive stars, it is most likely a YSO
(nevertheless, since it does not formally meet our selection criteria
it is not included in our analysis).  If so, the fitted SEDs suggest
that this is a Stage I object with a luminosity of $3.3\times
10^{4}$~L$_{\odot}$ and a mass of 14.7~\msun.  However, it is
important to remember that because of the very high source density
here we may actually be seeing multiple unresolved YSOs.  In that
case, we would be likely to overestimate the mass of the most massive
YSO and underestimate the total mass of YSOs contained in this source.
This type of source confusion should not have a strong effect on the
SED fitting results, because the observed SED will be dominated by
that of the most massive embedded YSO.  A mid-IR spectrum of this
object is displayed in \citet[][peak C]{contursi00}.  Another bright
\emph{Spitzer} source, SSTS3MC~14.7748-72.1749, is located within
3\arcsec\ of Sc 1 and has a steeply-rising mid- and far-IR SED that we
are unable to fit successfully.  If we remove the 24~\um\ upper limit
and either the 70~\um\ or $J$- and $K$-band detections (perhaps
justified because of confusion in this very densely populated region),
then this source has the SED of a very massive ($M > 10$~\msun),
early-stage YSO.  We do not detect any non-stellar sources in Sc~3
(immediately south of Sc~1), but confusion as a result of the very
bright source just north of the cluster may play a role in this
non-detection.

Sc~2 is located very close to 2 \emph{Spitzer} sources.
SSTS3MC~14.7574 has a stellar SED with a strong 24~\um\ detection
(indicating either confusion or circumstellar dust) and
SSTS3MC~14.7580-72.1763 is formally a poor fit as a result of a low
$K$-band flux and slight PAH contamination, but appears to have an SED
consistent with being an early stage YSO.

Scs 4-6 each have bright infrared counterparts.  Sc~4
is coincident with SSTS3MC~14.7605-72.1687, a massive embedded YSO.
Sc~5 is SSTS3MC~14.7514-72.1681, another object that can be
fitted by both YSO and AGB SEDs.  Again, given its position in a very
young cluster, it is most likely a massive stage I YSO rather than an
evolved star.  And Sc~6 contains 2 blended \emph{Spitzer}
sources, SSTS3MC~14.7371-72.1651 and SSTS3MC~14.7380-72.1651, both
very high-mass stage I YSOs.  The caveat mentioned above about
multiple unresolved sources applies here as well.  

Located on the southern edge of Sc~7, very close to the
prominent dust lane that arcs nearly halfway around N66, is
SSTS3MC~14.7733-72.1835, yet another very massive (18.6~\msun) stage I
protostar that can also be fit by AGB models.  Slightly further to the
southwest are 2 additional, lower-mass YSOs, SSTS3MC~14.7673-72.1834
and SSTS3MC~14.7698-72.1846.

Sc~8 has a faint \emph{Spitzer} counterpart, SSTS3MC~14.7816-72.1802,
which shows a stellar SED in the optical and near-IR and excesses in
the IRAC bands, but is not fit well by any of our YSO models.  It is
possible that the optical and IR emission is coming from different
sources.

Scs 9-11 are located around a quintet of YSOs.  The 2
brightest of these, SSTS3MC~14.8112-72.1843 and
SSTS3MC~14.8130-72.1840, lie within Sc~10, but it is not clear
whether these are truly blended YSOs as opposed to a bright, extended
clump of dust.  The other 3 sources classified as YSOs in this region
are SSTS3MC~14.8041-72.1867, SSTS3MC~14.8068-72.1856, and
SSTS3MC~14.8174-72.1866.

Sc~12 contains two bright \emph{Spitzer} sources
(SSTS3MC~14.8318-72.1890 and SSTS3MC~14.8353-72.1892) that are not fit
well by any of the available models.  The rising SEDs of these objects
towards long wavelengths strongly suggest that there are YSOs present
here, but blending may be a problem.  This cluster is coincident with
the source labeled peak I by \citet{contursi00}.

Scs 13-15 lie in the northward extension of the \hii\ region (N66A).
Clusters 13 and 14 each overlap with a YSO (SSTS3MC~14.8009-72.1663 =
\citeauthor{contursi00} peak F and SSTS3MC~14.8080-72.1578 =
\citeauthor{contursi00} peak G), but as with Sc~10 the source in Sc~14
does not appear pointlike in the IRAC bands, and so may not be a
single object.  Sc~15 is associated with SSTS3MC~14.8205-72.1544,
which shows very strong PAH emission but is not well-fit by YSO
models.

Finally, Sc~16 contains 2 \emph{Spitzer} sources, but both of these
are well-fit by normal stellar models, consistent with the older age
of this cluster derived by \citet{sabbi07}.

Out of the 44 YSOs in our sample that lie within the ACS \ha\ images
of \citet{nota06}, $\sim10$ are spatially coincident with gas or dust
pillars strongly resembling those made famous by \emph{HST} imaging of
the ``Pillars of Creation'' in the Eagle Nebula \citep{hester96}.
Higher spatial resolution near-IR and mid-IR imaging of these objects
may reveal exactly how the YSOs are related to these features.

\section{SUMMARY AND CONCLUSIONS}
\label{conclusions}

We have obtained mid-infrared imaging of the \hii\ region N66
(NGC~346) in the SMC with the IRAC and MIPS instruments on the
\emph{Spitzer Space Telescope}.  We detected 8011 unique sources, with
photometric coverage extending from $V$-band to 24~\um\ (70~\um\ in a
few cases).  Most of these sources have colors and SEDs consistent
with being normal stars, but we also detect a significant population
of objects that are very red in the mid-IR.  SED fitting of the 1645
sources with photometric measurements in at least four bands yielded
111 objects with SEDs that are best fit by YSO models rather than
stars or background galaxies.  These data represent the first
significant sample of embedded YSOs identified in an external galaxy.

We show that these YSOs can be mostly, but not completely, separated
from stars on the basis of their IRAC colors.  However, SED fitting is
necessary to significantly constrain the properties of individual
objects.  We find that the YSO models of
\citet{whitney03a,whitney03b,whitney04b}, which were designed to
represent Milky Way YSOs with solar metallicities, fit most of the
YSOs in N66 well, despite its much lower metallicity and dust-to-gas
ratio.  These results suggest that if low metallicity causes
significant changes in the star formation process, the threshold for
those effects must lie below the metallicity of the SMC (12 + log[O/H]
$\approx$ 8.0; \citealt{dufour75}).  The one possible difference
between SMC and Milky Way YSOs is that the SMC appears to contain a
population of very luminous but cool objects that have not been seen
in the Milky Way.  These objects may have expanded photospheres and
hence lower temperatures than normal because they are still accreting
material from their protostellar disks when they reach the main
sequence.  Alternatively, these sources could be multiple cool YSOs
that are unresolved by \emph{Spitzer}.
  
We calculate masses and stages (analogous to the usual YSO classes)
for each of the N66 YSOs, finding a range of masses from 2.4 to
16.6~M$_{\odot}$ (including 19 objects with masses above
8~M$_{\odot}$).  Almost half (45\%) of the YSOs are Stage II objects,
30\% are Stage I, and the remaining 25\% are evolved Stage III
sources.  We examine the spatial distribution of the YSOs and find
that they are strongly concentrated in the center of the \hii\ region
where bright \ha\ and dust emission is seen, but there are also small
numbers of YSOs in the surrounding region with much less diffuse ISM
emission.  The most massive YSOs are preferentially located closer to
the center of N66, indicating that mass segregation is taking place.
We compare our YSO map to the clusters of pre-main sequence stars
identified in the optical by \citet{sabbi07} and find that all but 2
of the clusters have associated YSOs.  Using a Salpeter IMF, we
calculate that a total of at least 3160~\msun\ of YSOs have been
formed in the last $\sim10^{6}$~yr, representing $\simgtr6$\% of the
current star formation in the SMC.

\acknowledgements
We would like to thank the anonymous referee for a
  careful reading of the paper that produced valuable feedback.
  J.D.S.  gratefully acknowledges the support of a Millikan Fellowship
  provided by the California Institute of Technology.  This research
  was partially funded by NASA through an award issued by JPL/Caltech
  (NASA-JPL \emph{Spitzer} grant 1264151 awarded to Cycle 1 project
  3316).  B.W. was supported by NASA Astrophysics Theory Program grant
  NNG05GH35G and the \emph{Spitzer} Theoretical Research Program under
  Subcontract 1290701, M.R. was supported by the Chilean {\sl Center
    for Astrophysics} FONDAP No. 15010003, and financial support from
  FONDECYT No. 1050052 is acknowledged by R.H.B.  We thank You-Hua
  Chu, Robert Gruendl, Lynne Hillenbrand, Jacco van Loon, Adam Leroy,
  and Bob Benjamin for helpful conversations, and we also thank
  Dimitrios Gouliermis for providing us with the \emph{HST} photometry
  in advance of publication.  This work is based on observations made
  with the \emph{Spitzer Space Telescope}, which is operated by the
  Jet Propulsion Laboratory, California Institute of Technology under
  a contract with NASA.  This publication makes use of data products
  from the Two Micron All Sky Survey, which is a joint project of the
  University of Massachusetts and the Infrared Processing and Analysis
  Center/California Institute of Technology, funded by NASA and the
  National Science Foundation.  This research has also made use of
  NASA's Astrophysics Data System Bibliographic Services and the
  SIMBAD database, operated at CDS, Strasbourg, France.

\clearpage
\LongTables
\begin{landscape}
\begin{deluxetable}{cccccccccccc}
\tabletypesize{\scriptsize}
\tablenum{1}
\tablewidth{0pt}
\tablecolumns{12}
\tablecaption{Young Stellar Objects in N66}
\tablehead{
\colhead{Number} & \colhead{Source name} & \colhead{$\alpha$ (J2000.0)}  & 
\colhead{$\delta$ (J2000.0)} & \colhead{$L$ (L$_{\odot}$)} &
\colhead{$\Delta L$\tablenotemark{a} (L$_{\odot}$)} & \colhead{$M_{*}$ (\msun)} & 
\colhead{$\Delta M_{*}$\tablenotemark{a} (\msun)} & \colhead{Stage} & \colhead{$\Delta$Stage\tablenotemark{a}} & 
\colhead{$\chi^{2}$ \tablenotemark{b}}  & \colhead{$\chi_{2}^{2}$ \tablenotemark{c}} \\
\colhead{(1)} & \colhead{(2)} & \colhead{(3)} & \colhead{(4)} &
\colhead{(5)} & \colhead{(6)} & \colhead{(7)} & \colhead{(8)} & 
\colhead{(9)} & \colhead{(10)} & \colhead{(11)} & \colhead{(12)} }
\startdata 
\cutinhead{Definite YSOs}
1   & SSTS3MC14.5039$-$72.0895 & 00 58 00.94 & $-$72 05 22.3 &  1290 &   556 &  6.5 & 0.8 & III & 0.6 & \phn0.26 & --- \\
2   & SSTS3MC14.5579$-$72.1554 & 00 58 13.89 & $-$72 09 19.5 &   131 &   --- &  5.1 & --- &  I  & --- & \phn1.76 & --- \\
3   & SSTS3MC14.6261$-$72.1449 & 00 58 30.26 & $-$72 08 41.5 &   328 &   107 &  4.6 & 0.4 & II  & 0.4 & \phn0.12 & --- \\
4   & SSTS3MC14.6680$-$72.1503 & 00 58 40.31 & $-$72 09 01.0 &  1590 &  1600 &  6.0 & 2.5 & II  & 0.5 & \phn0.40 & --- \\
5   & SSTS3MC14.7018$-$72.1834 & 00 58 48.43 & $-$72 11 00.2 &    35 &     2 &  2.8 & 0.5 &  I  & --- & \phn4.06 & --- \\
6   & SSTS3MC14.7094$-$72.1725 & 00 58 50.25 & $-$72 10 20.9 &  3300 &   --- &  8.3 & --- & III & --- & \phn5.53 & --- \\
7   & SSTS3MC14.7148$-$72.2427 & 00 58 51.55 & $-$72 14 33.9 &   671 &   353 &  5.3 & 0.9 & III & 0.5 & \phn0.95 & --- \\
8   & SSTS3MC14.7223$-$72.1679 & 00 58 53.35 & $-$72 10 04.3 &  3520 &   789 &  8.9 & 1.5 &  I  & --- & \phn1.39 & --- \\
9   & SSTS3MC14.7315$-$72.1734 & 00 58 55.56 & $-$72 10 24.3 &  4430 &   --- &  9.0 & --- & III & --- & \phn1.37 & --- \\
10  & SSTS3MC14.7329$-$72.2439 & 00 58 55.90 & $-$72 14 38.1 &   261 &   --- &  6.4 & --- &  I  & --- & \phn6.15 & --- \\
11  & SSTS3MC14.7429$-$72.1741 & 00 58 58.30 & $-$72 10 26.9 &  4340 &   --- &  7.9 & --- &  I  & --- & \phn3.55 & --- \\
12  & SSTS3MC14.7485$-$72.1843 & 00 58 59.63 & $-$72 11 03.3 &   348 &   --- &  4.5 & --- & II  & --- & \phn1.68 & --- \\
13  & SSTS3MC14.7509$-$72.1797 & 00 59 00.22 & $-$72 10 47.0 &  1940 &   --- &  7.3 & --- & III & --- & \phn2.37 & --- \\
14  & SSTS3MC14.7534$-$72.1829 & 00 59 00.81 & $-$72 10 58.6 &   388 &     1 &  4.7 & --- & II  & --- & \phn5.48 & --- \\
15  & SSTS3MC14.7566$-$72.1722 & 00 59 01.58 & $-$72 10 19.8 &  2230 &   --- &  7.5 & --- & II  & --- & \phn0.68 & --- \\
16  & SSTS3MC14.7582$-$72.2715 & 00 59 01.96 & $-$72 16 17.5 &   306 &   813 &  4.6 & 1.1 & II  & 0.7 & \phn0.41 & --- \\
17  & SSTS3MC14.7605$-$72.1687 & 00 59 02.53 & $-$72 10 07.4 &  5120 &  2210 & 10.3 & 0.8 &  I  & 0.4 & \phn3.69 & --- \\
18  & SSTS3MC14.7639$-$72.2068 & 00 59 03.34 & $-$72 12 24.3 &   197 &    46 &  3.9 & 0.2 & II  & --- & \phn0.56 & --- \\
19  & SSTS3MC14.7698$-$72.1846 & 00 59 04.75 & $-$72 11 04.7 &   684 &     7 &  5.5 & --- & II  & --- & \phn0.50 & --- \\
20  & SSTS3MC14.7766$-$72.2300 & 00 59 06.37 & $-$72 13 48.0 &   369 &   214 &  4.5 & 0.7 & II  & 0.5 & \phn0.00 & --- \\
21  & SSTS3MC14.7800$-$72.1454 & 00 59 07.19 & $-$72 08 43.4 &   261 &   --- &  6.4 & --- &  I  & --- & \phn9.19 & --- \\
22  & SSTS3MC14.7829$-$72.2091 & 00 59 07.90 & $-$72 12 32.9 &   459 &   151 &  4.9 & 0.4 & II  & --- & \phn6.59 & --- \\
23  & SSTS3MC14.7876$-$72.1769 & 00 59 09.03 & $-$72 10 37.0 &  4430 &   --- &  9.0 & --- & III & --- & \phn1.44 & --- \\
24  & SSTS3MC14.7880$-$72.1864 & 00 59 09.13 & $-$72 11 11.0 &  9880 &   --- & 11.6 & --- & III & --- & 13.16 & --- \\
25  & SSTS3MC14.7886$-$72.1827 & 00 59 09.25 & $-$72 10 57.8 & 30300 &   --- & 16.6 & --- &  I  & --- & \phn6.79 & --- \\
26  & SSTS3MC14.7898$-$72.1777 & 00 59 09.55 & $-$72 10 39.6 &  1620 &   --- &  7.0 & --- & III & --- & \phn3.99 & --- \\
27  & SSTS3MC14.7933$-$72.1930 & 00 59 10.40 & $-$72 11 34.9 &  6540 & 12500 &  7.9 & 8.3 &  I  & --- & 12.36 & --- \\
28  & SSTS3MC14.8009$-$72.1663 & 00 59 12.22 & $-$72 09 58.8 &  5060 &   --- & 10.1 & --- &  I  & --- & \phn9.19 & --- \\
29  & SSTS3MC14.8041$-$72.1867 & 00 59 12.98 & $-$72 11 12.3 &   176 &    15 &  3.8 & --- & II  & --- & \phn0.81 & --- \\
30  & SSTS3MC14.8056$-$72.2720 & 00 59 13.33 & $-$72 16 19.1 &   456 &   245 &  4.8 & 0.7 & III & 0.5 & \phn0.01 & --- \\
31  & SSTS3MC14.8061$-$72.1918 & 00 59 13.46 & $-$72 11 30.5 &  5990 &  2100 & 13.1 & 3.1 &  I  & --- & \phn6.52 & --- \\
32  & SSTS3MC14.8068$-$72.1856 & 00 59 13.62 & $-$72 11 08.0 &   540 &   131 &  5.2 & 0.4 & II  & --- & \phn0.18 & --- \\
33  & SSTS3MC14.8077$-$72.1466 & 00 59 13.85 & $-$72 08 47.7 &   728 &   159 &  5.6 & 0.3 & III & --- & \phn5.52 & --- \\
34  & SSTS3MC14.8080$-$72.1578 & 00 59 13.92 & $-$72 09 27.9 &  8380 &  1470 & 11.0 & 0.6 & II  & 0.7 & \phn8.32 & --- \\
35  & SSTS3MC14.8104$-$72.1621 & 00 59 14.49 & $-$72 09 43.7 &   387 &   --- &  4.7 & --- & II  & --- & \phn7.82 & --- \\
36  & SSTS3MC14.8112$-$72.1843 & 00 59 14.68 & $-$72 11 03.4 & 13000 &  7940 & 15.1 & 2.9 &  I  & --- & \phn1.17 & --- \\
37  & SSTS3MC14.8130$-$72.1840 & 00 59 15.11 & $-$72 11 02.2 &  9560 &  2180 & 11.4 & 0.8 & II  & 0.5 & \phn5.23 & --- \\
38  & SSTS3MC14.8140$-$72.1917 & 00 59 15.36 & $-$72 11 30.0 &  3300 &   --- &  8.3 & --- & III & --- & \phn0.42 & --- \\
39  & SSTS3MC14.8174$-$72.1866 & 00 59 16.18 & $-$72 11 11.6 &   210 &    61 &  4.0 & 0.2 & II  & --- & \phn1.00 & --- \\
40  & SSTS3MC14.8211$-$72.1822 & 00 59 17.07 & $-$72 10 55.8 &    87 &    10 &  4.5 & 0.3 &  I  & --- & \phn6.43 & --- \\
41  & SSTS3MC14.8284$-$72.1568 & 00 59 18.82 & $-$72 09 24.3 &  1180 &   --- &  6.4 & --- & II  & --- & \phn4.08 & --- \\
42  & SSTS3MC14.8292$-$72.1585 & 00 59 19.00 & $-$72 09 30.5 &   656 &   --- &  5.5 & --- & II  & --- & \phn4.88 & --- \\
43  & SSTS3MC14.8302$-$72.2441 & 00 59 19.25 & $-$72 14 38.8 &   453 &   174 &  4.9 & 0.8 & III & 0.5 & \phn0.00 & --- \\
44  & SSTS3MC14.8373$-$72.1449 & 00 59 20.95 & $-$72 08 41.5 &   922 &   345 &  6.2 & 0.4 & II  & 0.3 & \phn0.15 & --- \\
45  & SSTS3MC14.8465$-$72.1468 & 00 59 23.16 & $-$72 08 48.4 &  4430 &  1780 &  8.8 & 1.7 & III & 0.3 & \phn2.67 & --- \\
46  & SSTS3MC14.8487$-$72.1464 & 00 59 23.69 & $-$72 08 47.1 &   204 &    60 &  3.9 & 0.3 & II  & --- & \phn2.78 & --- \\
47  & SSTS3MC14.8551$-$72.1345 & 00 59 25.22 & $-$72 08 04.1 &  1340 &   --- &  6.6 & --- & III & --- & \phn0.40 & --- \\
48  & SSTS3MC14.8701$-$72.2089 & 00 59 28.83 & $-$72 12 31.9 &   131 &   --- &  5.1 & --- &  I  & --- & \phn6.78 & --- \\
49  & SSTS3MC14.8830$-$72.2265 & 00 59 31.91 & $-$72 13 35.3 &   206 &   --- &  5.9 & --- &  I  & --- & \phn5.28 & --- \\
50  & SSTS3MC14.8900$-$72.2789 & 00 59 33.61 & $-$72 16 43.9 &   678 &   215 &  5.5 & 0.5 & III & --- & \phn2.11 & --- \\
51  & SSTS3MC14.9039$-$72.1509 & 00 59 36.94 & $-$72 09 03.2 &    77 &    11 &  4.6 & 0.3 &  I  & 0.4 & \phn3.33 & --- \\
52  & SSTS3MC14.9041$-$72.2398 & 00 59 36.98 & $-$72 14 23.2 &  1000 &   495 &  5.8 & 1.5 & II  & 0.2 & \phn0.22 & --- \\
53  & SSTS3MC14.9360$-$72.2122 & 00 59 44.64 & $-$72 12 44.0 &  2150 &   --- &  7.5 & --- & III & --- & \phn7.94 & --- \\
54  & SSTS3MC14.9603$-$72.2764 & 00 59 50.48 & $-$72 16 35.2 &   644 &   249 &  5.4 & 0.6 & III & 0.5 & \phn0.51 & --- \\
55  & SSTS3MC14.9671$-$72.2613 & 00 59 52.12 & $-$72 15 40.8 &   529 &   376 &  5.0 & 0.9 & II  & 0.5 & \phn0.21 & --- \\
56  & SSTS3MC14.9858$-$72.2454 & 00 59 56.59 & $-$72 14 43.5 &   683 &   177 &  5.9 & 0.8 & II  & 0.4 & 11.79 & --- \\
57  & SSTS3MC15.0636$-$72.2746 & 01 00 15.27 & $-$72 16 28.7 &  1000 &   --- &  6.2 & --- & III & --- & \phn5.99 & --- \\
58  & SSTS3MC15.0655$-$72.0788 & 01 00 15.72 & $-$72 04 43.6 &   360 &    87 &  4.6 & 0.4 & II  & --- & \phn0.26 & --- \\
59  & SSTS3MC15.0859$-$72.2232 & 01 00 20.62 & $-$72 13 23.6 &   488 &   119 &  5.0 & 0.3 & II  & 0.5 & \phn7.91 & --- \\
60  & SSTS3MC15.0930$-$72.1662 & 01 00 22.32 & $-$72 09 58.2 &  2910 &   944 &  8.0 & 0.7 &  I  & 0.5 & \phn0.40 & --- \\
61  & SSTS3MC15.1415$-$72.2574 & 01 00 33.97 & $-$72 15 26.5 &  1750 &    99 &  8.8 & 0.4 &  I  & --- & \phn5.55 & --- \\
\cutinhead{Probable YSOs}
62  & SSTS3MC14.4895$-$72.2499 & 00 57 57.49 & $-$72 14 59.7 & 12700 & 10900 & 11.4 & 5.3 & III & --- & \phn1.95 & \phn3.49 \\
63  & SSTS3MC14.5083$-$72.0835 & 00 58 02.00 & $-$72 05 00.7 &  2000 &  2130 &  7.0 & 2.3 & II  & 0.4 & \phn0.04 & \phn8.60 \\
64  & SSTS3MC14.5104$-$72.2020 & 00 58 02.50 & $-$72 12 07.0 &   573 &   536 &  5.1 & 1.5 & II  & 0.8 & \phn0.02 & \phn4.23 \\
65  & SSTS3MC14.5410$-$72.1448 & 00 58 09.85 & $-$72 08 41.2 &   160 &   107 &  4.0 & 0.8 & II  & 0.5 & \phn0.28 & \phn1.45 \\
66  & SSTS3MC14.5911$-$72.1983 & 00 58 21.87 & $-$72 11 53.8 &   114 &   --- &  5.0 & --- &  I  & --- & \phn5.85 & 10.50 \\
67  & SSTS3MC14.5959$-$72.2472 & 00 58 23.01 & $-$72 14 50.0 &    95 &    93 &  4.3 & 0.9 &  I  & 0.4 & \phn0.60 & \phn7.06 \\
68  & SSTS3MC14.6593$-$72.2436 & 00 58 38.23 & $-$72 14 37.0 &   210 &   300 &  3.7 & 1.1 & II  & 0.5 & \phn0.01 & \phn2.29 \\
69  & SSTS3MC14.6977$-$72.1509 & 00 58 47.45 & $-$72 09 03.2 &  2290 &   188 &  7.6 & 0.2 & III & --- & \phn5.99 & 14.70 \\
70  & SSTS3MC14.7067$-$72.1161 & 00 58 49.60 & $-$72 06 58.0 &  1470 &  2230 &  6.5 & 1.2 & II  & 0.3 & \phn0.05 & \phn3.93 \\
71  & SSTS3MC14.7169$-$72.1788 & 00 58 52.04 & $-$72 10 43.8 &   517 &   900 &  4.8 & 0.9 & II  & 0.2 & \phn2.20 & \phn7.18 \\
72  & SSTS3MC14.7173$-$72.1293 & 00 58 52.16 & $-$72 07 45.4 &   260 &   317 &  4.3 & 0.6 & II  & 0.2 & \phn0.07 & \phn1.78 \\
73  & SSTS3MC14.7175$-$72.2081 & 00 58 52.19 & $-$72 12 29.1 &    80 &    16 &  4.1 & 1.0 &  I  & --- & \phn4.49 & 12.55 \\
74  & SSTS3MC14.7177$-$72.2063 & 00 58 52.24 & $-$72 12 22.6 &    37 &    69 &  2.9 & 0.8 & II  & 0.6 & \phn0.00 & \phn0.32 \\
75  & SSTS3MC14.7208$-$72.1536 & 00 58 53.00 & $-$72 09 12.9 &  1640 &   307 &  7.0 & 0.4 & III & --- & \phn0.06 & \phn1.01 \\
76  & SSTS3MC14.7265$-$72.1957 & 00 58 54.36 & $-$72 11 44.6 &    59 &   105 &  2.4 & 4.6 &  I  & --- & \phn0.01 & \phn2.38 \\
77  & SSTS3MC14.7269$-$72.2737 & 00 58 54.45 & $-$72 16 25.1 &  1170 &   861 &  6.4 & 2.3 & III & 0.5 & \phn0.00 & \phn2.29 \\
78  & SSTS3MC14.7327$-$72.1420 & 00 58 55.85 & $-$72 08 31.3 &   236 &   --- &  4.0 & --- &  I  & --- & \phn3.05 & 11.44 \\
79  & SSTS3MC14.7371$-$72.1651 & 00 58 56.89 & $-$72 09 54.5 &  9390 &  3200 & 13.5 & 1.8 &  I  & --- & \phn3.54 & 10.49 \\
80  & SSTS3MC14.7380$-$72.1651 & 00 58 57.11 & $-$72 09 54.3 & 14800 &  7190 & 15.5 & 2.4 &  I  & --- & \phn0.18 & \phn1.96 \\
81  & SSTS3MC14.7383$-$72.1449 & 00 58 57.19 & $-$72 08 41.5 &   231 &   370 &  4.1 & 1.0 & II  & 0.5 & \phn0.02 & \phn1.76 \\
82  & SSTS3MC14.7407$-$72.2761 & 00 58 57.76 & $-$72 16 34.0 &  1210 &   874 &  6.4 & 0.6 & II  & 0.6 & \phn0.48 & \phn1.57 \\
83  & SSTS3MC14.7426$-$72.1701 & 00 58 58.23 & $-$72 10 12.4 &  3290 &   --- &  8.3 & --- & III & --- & \phn0.32 & \phn1.64 \\
84  & SSTS3MC14.7542$-$72.2250 & 00 59 01.01 & $-$72 13 30.1 &   257 &   469 &  4.2 & 0.8 & II  & 0.4 & \phn0.00 & \phn3.06 \\
85  & SSTS3MC14.7569$-$72.2725 & 00 59 01.65 & $-$72 16 21.1 &   251 &   --- &  6.5 & --- &  I  & --- & \phn2.77 & \phn8.52 \\
86  & SSTS3MC14.7577$-$72.2715 & 00 59 01.86 & $-$72 16 17.4 &   726 &  1020 &  5.2 & 1.4 & II  & 0.6 & \phn0.00 & \phn6.20 \\
87  & SSTS3MC14.7673$-$72.1834 & 00 59 04.14 & $-$72 11 00.1 &  2290 &  1870 &  7.0 & 1.9 & II  & 0.7 & \phn2.42 & \phn8.66 \\
88  & SSTS3MC14.7702$-$72.2090 & 00 59 04.84 & $-$72 12 32.5 &   310 &    47 &  6.7 & 0.5 &  I  & --- & \phn0.96 & \phn3.43 \\
89  & SSTS3MC14.7814$-$72.1369 & 00 59 07.54 & $-$72 08 12.8 &   176 &   151 &  3.6 & 0.7 & III & 0.5 & \phn0.50 & \phn2.90 \\
90  & SSTS3MC14.7872$-$72.1781 & 00 59 08.93 & $-$72 10 41.2 &  2360 &   --- &  7.6 & --- & II  & --- & \phn0.20 & \phn6.33 \\
91  & SSTS3MC14.7882$-$72.1441 & 00 59 09.17 & $-$72 08 38.7 &   201 &   914 &  3.9 & 1.2 & II  & 0.5 & \phn0.12 & \phn1.09 \\
92  & SSTS3MC14.7893$-$72.1401 & 00 59 09.43 & $-$72 08 24.2 &    80 &    40 &  3.6 & 0.5 & II  & 0.5 & \phn0.10 & \phn7.14 \\
93  & SSTS3MC14.7900$-$72.0911 & 00 59 09.60 & $-$72 05 27.9 &   821 &   226 &  5.8 & 0.4 & III & 0.7 & \phn0.48 & \phn3.25 \\
94  & SSTS3MC14.7908$-$72.2089 & 00 59 09.80 & $-$72 12 31.9 &    57 &    23 &  2.7 & 0.2 & II  & --- & \phn0.38 & \phn8.03 \\
95  & SSTS3MC14.7909$-$72.1970 & 00 59 09.81 & $-$72 11 49.1 &  2270 &  1610 &  7.2 & 1.7 & III & 0.5 & \phn0.49 & \phn9.59 \\
96  & SSTS3MC14.7911$-$72.2052 & 00 59 09.87 & $-$72 12 18.8 &   162 &    74 &  3.7 & 0.3 & II  & 0.3 & \phn0.05 & \phn6.49 \\
97  & SSTS3MC14.7956$-$72.1865 & 00 59 10.93 & $-$72 11 11.5 &   114 &   373 &  3.2 & 1.5 & II  & --- & \phn0.01 & \phn3.12 \\
98  & SSTS3MC14.7988$-$72.1372 & 00 59 11.72 & $-$72 08 13.7 &   114 &   --- &  5.0 & --- &  I  & --- & \phn6.32 & 12.05 \\
99  & SSTS3MC14.8006$-$72.1939 & 00 59 12.13 & $-$72 11 38.1 &   171 &    54 &  4.1 & 0.7 &  I  & 0.5 & \phn1.69 & \phn8.27 \\
100 & SSTS3MC14.8011$-$72.1935 & 00 59 12.25 & $-$72 11 36.7 &   305 &   228 &  4.2 & 0.8 & II  & --- & \phn0.07 & \phn7.34 \\
101 & SSTS3MC14.8426$-$72.2461 & 00 59 22.23 & $-$72 14 45.8 &   275 &   170 &  4.7 & 0.6 & II  & 0.2 & \phn0.00 & \phn0.12 \\
102 & SSTS3MC14.8466$-$72.1513 & 00 59 23.19 & $-$72 09 04.5 &   356 &   109 &  4.5 & 0.6 & II  & --- & \phn2.22 & \phn6.76 \\
103 & SSTS3MC14.8532$-$72.1669 & 00 59 24.77 & $-$72 10 00.8 &  1160 &    90 &  7.0 & 0.7 & II  & 1.0 & \phn0.70 & \phn7.04 \\
104 & SSTS3MC14.8715$-$72.1774 & 00 59 29.15 & $-$72 10 38.6 &   490 &    58 &  7.3 & 0.3 &  I  & --- & \phn2.46 & \phn5.53 \\
105 & SSTS3MC14.8729$-$72.1393 & 00 59 29.49 & $-$72 08 21.6 &   123 &    45 &  4.4 & 0.8 &  I  & --- & \phn2.10 & \phn4.92 \\
106 & SSTS3MC14.8781$-$72.1694 & 00 59 30.74 & $-$72 10 09.9 &   441 &   385 &  4.4 & 1.5 & III & 0.5 & \phn0.97 & \phn9.78 \\
107 & SSTS3MC14.8817$-$72.2385 & 00 59 31.60 & $-$72 14 18.6 &   540 &   963 &  4.6 & 1.6 & II  & 0.7 & \phn0.16 & \phn6.91 \\
108 & SSTS3MC14.8848$-$72.0817 & 00 59 32.35 & $-$72 04 54.2 &    98 &    47 &  4.2 & 1.1 &  I  & --- & \phn1.85 & \phn6.99 \\
109 & SSTS3MC14.9738$-$72.1101 & 00 59 53.70 & $-$72 06 36.3 &   201 &   215 &  4.3 & 0.7 & II  & 0.6 & \phn0.05 & \phn4.07 \\
110 & SSTS3MC15.0896$-$72.1743 & 01 00 21.49 & $-$72 10 27.5 &    80 &    30 &  3.9 & 0.9 &  I  & 0.5 & \phn0.14 & \phn3.10 \\
111 & SSTS3MC15.1199$-$72.1377 & 01 00 28.78 & $-$72 08 15.6 &   847 &   166 &  5.9 & 0.3 & III & --- & \phn0.06 & \phn6.27 \\
\enddata
\tablecomments{The coordinates in this table are not necessarily
identical to the ones presented by B07 and made available on the
S$^{3}$MC website because the N66 photometry was carried out
independently.  However, the differences should be very small.}

\tablenotetext{a}{The uncertainties on the luminosities, masses, and
  stages are calculated as the weighted standard deviation of the
  luminosities, masses, and stages of all of the acceptable YSO
  models.  In cases where there is only one acceptable model for a
  given source, or there are multiple models but they all produce the
  same luminosity/mass/stage, we cannot calculate an uncertainty in
  this way, so we leave the corresponding space in the table blank.}

\tablenotetext{b}{$\chi^{2}$ value for the best fitting SED.  Note
  that these are raw $\chi^{2}$ values, not $\chi^{2}$ per data point,
  so a good fit ($\chi^{2}$ per data point $ \le 2.2$) can have a
  $\chi^{2}$ as high as 8.8 (if there are 4 flux data points for that
  source) or more.}
\tablenotetext{c}{$\chi^{2}$ value for the best fitting AGB star or
  background galaxy SED.}
\label{ysotable}
\end{deluxetable}
\clearpage
\end{landscape}

\begin{deluxetable}{lc}
\tablenum{2}
\tablecolumns{2}
\tablecaption{SED Classification Results}
\tablehead{
\colhead{\hspace{-1.22in}Object Class} & \colhead{Number} }

\startdata 
All sources & 1645 \\
\cutinhead{Objects with unique SED classifications}
Normal stars & 1322 \\
YSOs & 61 \\
Galaxies & 27 \\
AGB stars & 6 \\
\cutinhead{Objects without unique SED classifications}
Probable YSOs & 50 \\
Probable galaxies & 13 \\
Probable AGB stars & 5 \\
Multiple good fits; unable to classify & 13 \\
Poor fits & 148
\enddata
\label{sourcetypes}
\end{deluxetable}

\end{document}